\documentclass{IEEEtran}

\ifCLASSINFOpdf
\usepackage[pdftex]{graphicx}
\else
\usepackage[dvips]{graphicx}
\fi
\usepackage[cmex10]{amsmath}
\usepackage{amsfonts}
\usepackage{acronym}
\usepackage{color}
\usepackage{comment}
\usepackage{cite}

\usepackage{amsthm}

\theoremstyle{definition}
\newtheorem*{theorem*}{Theorem}

\begin{document}

\title{Decentralized Approach to Detect and Eliminate Flapping Phenomena due to Flexible Resources}

\author{\IEEEauthorblockN{Angel Vaca, Federico
    Milano\\} \IEEEauthorblockA{School of Electrical \& Electronic
    Engineering, University College Dublin (UCD), Ireland \\
    angel.vaca1@ucdconnect.ie, federico.milano@ucd.ie }}

\maketitle

\begin{abstract}
This paper presents a decentralized methodology for detecting and mitigating flapping phenomena in power systems, primarily caused by the operation of discrete devices.  The proposed approach applies moving-window autocorrelation to local measurements, enabling each device to autonomously identify sustained oscillations.  Upon detection, a probabilistic, device-specific mitigation strategy is executed.  Flexible demand resources (DFRs), under-load tap changers (ULTCs), and automatic voltage regulators (AVRs) are utilised to illustrate the performance of the proposed approach to both discrete and continuous-operation devices.  Results show that the proposed method is robust and properly distinguishes damped oscillations from persistent flapping, allowing devices to independently recognize problematic operating scenarios and implement corrective actions accordingly.

\end{abstract}


\vspace{-1mm}
\section{Introduction}

\vspace{-2mm}
\subsection{Motivation}

The modernization of power systems is characterized by an increasing proliferation of granular and discrete flexible resources (DFRs), such as distributed energy resources and demand response units.  These devices are characterized by their small individual capacities but extensive distribution throughout the network.  Typically, they operate discretely, switching between predefined operational states according to local measurement signals, primarily to support frequency or voltage stability.  Other, larger devices such as Under Load Tap Changers (ULTC) or static reactive power compensators have a similar operation logic, by following a signal from the system, and reacting by switching to a different discrete level.

While such control schemes of the above-mentioned devices offer operational flexibility, they can inadvertently introduce undesirable system-wide phenomena, as extensively documented in the literature \cite{1652983}.  For instance, studies such as \cite{challenges} highlight the potential issues arising from numerous discrete flexible resources (DFRs) responding independently to frequency deviations, provoking a so-called \textit{flapping} effect, characterized by sustained oscillatory behaviour due to frequent switching between operational states.  Similarly, the work presented in \cite{4349071} illustrates how multiple ULTCs operating sequentially along the same transmission path can engage in persistent and counterproductive switching cycles.

Finally, it is well known that certain continuous controllers, such as the automatic voltage regulators (AVRs) of synchronous machines can lead to limit cycles \cite{544628, 6062658}, which can be interpreted as the continuous counterpart of the flapping phenomenon.

This work proposes a technique based on the estimation of the autocorrelation of measured signals to clear the oscillating phenomena caused by discrete and continuous controller shortly after their occurrence.   This approach allows fully exploiting the capability of devices equipped with discrete or continuous control, while limiting the effects of arising persistent oscillations. 

\subsection{Literature Review}

The phenomenon of flapping—sustained, undesired oscillations arising from uncoordinated switching of discrete devices—has been extensively documented.  Early analyses in \cite{1652983} and \cite{hiskens_trajectory_2000} characterize stability regions for load tap changers and highlight the risk of limit cycles under certain regulator deadband settings.  Grazing bifurcations in cascaded ULTCs were studied in \cite{4349071}, while \cite{Gubner_2006} demonstrates how interactions among DFRs and ULTCs can provoke counter‐productive switching cycles.  Boundary-value approaches, as cited in \cite{1525127}, demonstrate that nonsmooth device dynamics result in complex cyclic behavior, with stability being critically dependent on system inertia and switching thresholds.

A range of mitigation strategies has been proposed.  Dead-bands and rate limiters reduce device activity at the cost of responsiveness \cite{challenges}.  Others, such as partial centralization of control, utilization of deadbands, and adjustments in stochastic controls, have been proposed to mitigate the cycling effects but require aspects as communication and continuous adjustments of parameters \cite{challenges, microflexibility,nayyar_decentralized_2013}.  Work on DFRs for ancillary services has demonstrated their value in frequency regulation \cite{kirby1999load,lu2006design,short2007stabilization}, yet also underscores the persistent risk of induced oscillations when operating stochastically and independently \cite{heffner2008loads}.

Autocorrelation analysis provides a powerful, noise-robust means to detect periodic or seasonal behavior and has seen broad application across various engineering fields \cite{10783240}, pitch extraction in speech processing \cite{952490}, and multipath error mitigation in communication systems \cite{570787}.  In power systems, exponentially decaying autocorrelation models underpin wind‐speed stochastic differential equations \cite{ZARATEMINANO2016186,7741754,jonsdottir_data-based_2019}, and recent work shows that load autocorrelation influences voltage stability margins \cite{9637935}.  Based on some relevant features of the autocorrelation, we develop an autocorrelation-based detection framework, along with complementary self-actions for the devices to control flapping.

\subsection{Contributions}
This paper makes two main contributions:
\begin{itemize}
  \item A decentralized detection algorithm based on signal autocorrelation that identifies persistent flapping in discrete devices using only local measurements.
  \item Design of adaptive control logic for DFRs, ULTCs, and continuous elements as AVRs to autonomously mitigate flapping, preserving device flexibility while enhancing system stability.
\end{itemize}

\subsection{Paper Organization}
The remainder of the paper is organized as follows.  Section II presents the theoretical foundations and derivation of the autocorrelation-based flapping detection methodology and the devices' models.  Section III describes case studies, analyzing the application of the model to DFRs, ULTCs, and continuous-nature operation AVRs.  Section IV proposes practical considerations, including parameter selection and robustness to noise and model uncertainty.  Finally, Section V concludes the paper and outlines directions for future research.

\section{Autocorrelation-Based Flapping Detection}
\label{sec:method}

As mentioned in the context of discrete devices, undesired oscillatory behavior can arise through different mechanisms.  The flapping effect denotes sustained oscillations driven by uncoordinated or counter-productive switching actions, while limit cycles refer to self-sustained oscillations inherent to the system dynamics.  For the purposes of this work, the term flapping effect will be used to describe the sustained oscillatory patterns of interest, regardless of whether they originate from control-driven dynamics or hybrid switching mechanisms.

Autocorrelation analysis provides a principled approach to quantify repetitive patterns in a time series and to distinguish persistent flapping from benign transients.  In this work, these characteristics are exploited to develop a method capable of reliably detecting flapping events directly from device measurements.  In this section, we develop the theoretical foundations, explain each extension to the autocorrelation formulation, and justify the design choices that yield a robust, on‐device detection algorithm.

\subsection{Notation}
The following notation is utilized:

\begin{itemize}
  \item $f_{i,k}$: raw measurement (frequency or voltage) of device $i$ at sample $k$.
  \item $\Delta t$: sampling interval (s).
  \item $N_w,N_s$: window length and shift (samples).
  \item $\mu_{i,n},\sigma_{i,n}$: windowed mean and standard deviation.
  \item $\hat R[k],\rho[k]$: biased sample autocorrelation and normalized coefficient.
  \item $T_{\min},T_{\max}$: expected flapping period bounds (s).
  \item $k_{\min},k_{\max}$: corresponding lag indices.
  \item $r_i^*$: flapping detection metric.
  \item $R_{\mathrm{th}}$: autocorrelation threshold.
  \item $\varepsilon$: tolerance on metric decay.
  \item $M$: required consecutive detections.
  \item $c_i$, $\phi_i$: persistence counter and flapping flag.
\end{itemize}

\subsection{Continuous and Sample Autocorrelation}
For a real, wide–sense stationary process $X(t)$, the autocorrelation function
\begin{equation}
R_X(\tau) = \mathbb{E}\bigl[X(t)\,X(t+\tau)\bigr]
\label{eq:acf_continuous_expanded}
\end{equation}
measures the expected similarity of the signal with itself delayed by $\tau$.  Because of stationarity, $R_X(\tau)$ depends only on the lag $\tau$, not on the absolute time $t$.  Thus, for a sequence $x[n]$, we define the biased sample autocorrelation:
\begin{equation}
\hat R[k] \;=\; \frac{1}{N}\sum_{n=0}^{N - k - 1} x[n]\,x[n + k],
\quad k = 0,1,\dots,N-1.
\label{eq:acf_sample_expanded}
\end{equation}
Dividing by $N$ simplifies the statistical bias analysis and ensures that all lags are comparable.  We then normalize:
\begin{equation}
\rho[k] = \frac{\hat R[k]}{\hat R[0]},
\quad
\rho[0] = 1,\quad
|\rho[k]| \le 1.
\label{eq:acf_norm_expanded}
\end{equation}
The normalized autocorrelation $\rho[k]$ captures only temporal structure, independent of signal amplitude.

\subsection{Windowing: Time-Local Analysis}
Real-world frequency or voltage measurements are nonstationary: their statistical properties drift over time.  To focus on recent dynamics, we apply a moving window of length $N_w$ samples ($T_w = N_w\Delta t$), shifting by $N_s$ samples ($T_s = N_s\Delta t$).  At each analysis step $n$, device $i$ constructs
\[
\mathbf{f}_i^{(w)} = \bigl[f_{i,n-N_w+1},\ldots,f_{i,n}\bigr].
\]
Windowing adapts the detector to evolving operating conditions and bounds computational load, since $N_w$ is chosen to span the longest expected flapping period.

\subsection{Signal Normalization: Removing Bias and Scale}
Within each window, we remove any DC offset and rescale to unit variance:

\begin{equation}
\mu_{i,n} = \frac{1}{N_w}\sum_{m=1}^{N_w} f_{i,n-N_w+m},
\end{equation}

\begin{equation}
\sigma_{i,n}^2 = \frac{1}{N_w}\sum_{m=1}^{N_w}\bigl(f_{i,n-N_w+m}-\mu_{i,n}\bigr)^2,
\end{equation}

\begin{equation}
x_{i,m} = \frac{f_{i,n-N_w+m} - \mu_{i,n}}{\sigma_{i,n}}.
\end{equation}

Subtracting the mean ($\mu_i$) removes any DC component, isolating fluctuations around the typical value.  This ensures zero mean (DC bias removed) and unit variance (dimensionless comparison).  Complementary, dividing by the standard deviation ($\sigma_i$) standardizes the variance to 1, ensuring that only the temporal structure of the signal is analyzed, so the autocorrelation metric is not biased by varying signal amplitude.  This standardization step is critical to apply a single detection threshold $R_{\mathrm{th}}$ across different devices and operating scenarios.

\subsection{Lag-Band Selection: Focusing on Flapping Periods}
Flapping oscillations occur within a known period range $[T_{\min},T_{\max}]$.  We convert these bounds to integer lags:
\begin{equation}
k_{\min} = \bigl\lceil T_{\min}/\Delta t\bigr\rceil,
\quad
k_{\max} = \bigl\lfloor T_{\max}/\Delta t\bigr\rfloor.
\label{eq:lag_bounds_expanded}
\end{equation}
By restricting the autocorrelation analysis to $k\in[k_{\min},k_{\max}]$, we suppress high-frequency noise ($k<k_{\min}$) and slow drifts ($k>k_{\max}$), effectively creating a data-driven bandpass filter that isolates candidate flapping periodicities.

\subsection{Detection Metric and Persistence Criterion}
Within the selected lag band, the maximum absolute normalized autocorrelation
\begin{equation}
r_i^* = \max\bigl|\rho[k]\bigr|
\label{eq:metric_expanded}
\end{equation}
serves as a robust indicator of periodicity strength.  To guard against false alarms from transient spikes or random fluctuations, we require that $r_i^*$ exceed the threshold $R_{\mathrm{th}}$ for $M$ consecutive windows, allowing a small tolerance $\varepsilon$ for natural decay:
\begin{align}
c_i &=
\begin{cases}
c_i + 1, & r_i^* > R_{\mathrm{th}}
\ \wedge\ r_i^* \ge r_{i-1}^* - \varepsilon,\\
0, & \text{otherwise},
\end{cases}
\label{eq:counter_expanded}\\
\phi_i &=
\begin{cases}
1, & c_i \ge M,\\
0, & \text{otherwise}.
\end{cases}
\label{eq:flag_expanded}
\end{align}
Here, $c_i$ is a persistence counter, and $\phi_i=1$ flags confirmed flapping at device $i$.  The tolerance $\varepsilon$ accounts for small variations in autocorrelation due to window shifts.  The threshold $R_{th}$ can be chosen empirically or via statistical analysis of typical responses.

\section{Models and Response Actions of the Devices}
\label{sec:models}

Once flapping is detected, a range of corrective actions can be taken to restore system stability.  The specific response depends on the operating nature of each device.  In what follows, we describe the representative strategies adopted by granular, cascaded, and non-granular devices, highlighting both direct interventions and advanced coordination mechanisms.

\subsection{Stochastic Control}  

Stochastic control has been proposed for granular devices that form large populations of small, homogeneous units distributed across the system.  These devices act independently, without communication, but follow similar control logic.

We model each DFR $i$ as injecting power 
\begin{equation} p_i(t) = \kappa_i(t)\,p_{i,o}, 
\end{equation} 
where $p_{i,o}$ is the constant power step size and $\kappa_i(t)\in\{0,1\}$ its on/off state \cite{flexiLoads}.  Switching follows a decentralized stochastic policy evaluated every $\Delta t_{\rm ctrl}$: 
\begin{equation} 
u_i(t) = \frac{\Delta\omega(t) + \Delta\omega_{\mathrm{thr}}}{2\,\Delta\omega_{\mathrm{thr}}}, 
\end{equation} 

\begin{equation} \beta_i(t) = \frac{n}{k(t)}\,\frac{m(t)}{m_{\mathrm{ref}}}, 
\end{equation} 
where $\Delta\omega$ is the local frequency deviation, $\Delta\omega_{\rm thr}$ a dead-band, $n$ the total DFR count, $k(t)$ the subset currently available, and $m(t)$ the estimated inertia with reference $m_{\rm ref}$.  Once the value of $u_i$ is determined, each DFR independently generates a random number between 1 and 0 using a uniform distribution, say $X \sim \mathcal{U}(0, 1)$.

Then the flexible load will switch on or off if: 
\begin{equation}
\label{eq:logicDFR} 
\kappa_i(t) = \begin{cases} 1, & X \le \beta_i(t)\,u_i(t),\\ 0, & \text{otherwise}.  \end{cases} 
\end{equation}
Some references, as \cite{UCTEhandbook, challenges}, propose that the coefficient $\beta_i$ is updated every hour or half an hour and made available by the system operator, making the system not entirely decentralized.  In the proposed method, we do not require central coordination that utilizes the number of DFRs or the current inertia of the system, $\beta_i(t)$.  Instead, $ \beta_i (t) $ is considered a factor that modulates the stochastic operation of the DFRs.  This model captures both the inherent randomness of large populations and the inertia‐aware modulation of switching probability, ensuring that fewer DFRs respond as system inertia decreases.

Upon detection of flapping, the decentralized flexible resource (DFR) $i$ updates its modulation factor according to  
\begin{equation}
\beta_i(t) = a \, \beta_{i,\text{init}}(t),
\end{equation}  
where $a \in (0,1)$.  This reduction lowers the switching probability and damps the oscillatory cycle.  Importantly, residual flexibility is preserved, enabling rapid restoration of support once stability is regained.
To have a more conservative action and block the source of oscillation, one can set up a complete disconnection of the devices, with $a = 0$.

\subsection{Cascading}  

Devices connected along the same feeder (cascade), are particularly prone to inducing flapping by oscillating against one another.  This phenomenon has been observed in particular for discrete controllers with large steps, such as the case of ULTCs connected in cascade.  In these cases, the most effective action is to block the operation of one device, thereby breaking the cycling loop and preserving the ability of the remaining unit(s) to regulate their target variable.  If oscillations persist, subsequent devices may also activate their blocking logic in sequence until the cycling is suppressed.

\subsection{Limit Cycles}

In contrast to stochastic control, large continuous devices such as synchronous machines with primary controllers can give rise to limit cycles due to the dynamic coupling of their dynamics under specific operating conditions, typically characterised by weak grid conditions.  Due to the large size of most conventional power plants, they cannot be disconnected.  To avoid limit cycles, the most common solution is to include power system stabilizers \cite{4110969}.  However, even with proper tuning, it can never be completely excluded the possibility that a limit cycle occurs.  We propose thus an alternative remedial action that combines stochastic control and the adjustment of the internal control parameters of the primary regulation associated with the oscillatory behaviour.  

To guide this coordination in cases of flapping detection, we propose the use of the Teager Energy Operator (TEO) \cite{chen_forced_2024, kamwa_robust_2011},
a nonlinear tool that estimates the instantaneous energy of a signal by combining amplitude and frequency information.  For a discrete-time signal $x[n]$, the operator is defined as
\begin{equation}
\Psi[x[n]] = x[n]^2 - x[n+1] \cdot x[n-1].
\end{equation}  

The TEO is particularly effective in detecting and quantifying oscillatory behavior in nonstationary signals.  On a flapping event, the first step is to extract the TEO of a representative feature that reflects the device’s contribution to the disturbance.  This feature may be defined in different ways, for example, as the slope of the ray connecting the oscillation’s starting point to its value after a given time, or as the angle ($\alpha_i$) of that ray measured from the reference (the $X$‐axis).  

Normalization by the maximum possible value yields a dimensionless metric
\begin{equation}
\label{eq:alfa}
\tilde{\alpha}_i = \frac{\alpha_i}{\alpha_{\text{max}}}, \quad \tilde{\alpha}_i \in [0,1],
\end{equation}
which provides a comparable measure of relative impact across devices.

Based on this metric, we propose a stochastic local decision-making scheme that preserves decentralization.  At every control interval $M_{\rm ctrl}$, device $i$ updates its parameter that causes the flapping as
\begin{equation}
\label{eq:TEOoperation}
K_{a,i} =
\begin{cases}
K_{a,i}^{\text{safe}}, & \text{if } X \leq \tilde{\alpha}_i,\\
K_{a,i}^{\text{init}}, & \text{otherwise},
\end{cases}
\end{equation}
where $X \sim \mathcal{U}(0,1)$.  Devices with greater estimated impact are thus more likely to switch to a safe mode, contributing proportionally to damping while maintaining full decentralization and avoiding coordination delays.

\section{Case Studies}
\label{sec:case}

The proposed autocorrelation-based detection framework is implemented in the power system software tool Dome \cite{vancouver} and tested on the WSCC 9-bus and IEEE 14-bus networks.  We design scenarios that deliberately provoke flapping by high-activity switching, considering three representative device classes: Discrete Flexible Resources (DFRs) and Under-Load Tap Changers (ULTCs) as discrete devices, and Automatic Voltage Regulators (AVRs) as devices with continuous-nature operation.  These case studies allow us to assess detection accuracy, response latency, and the trade-offs between flexibility and stability under different loading and inertia conditions.

\subsection{Discrete Flexible Resources}
\label{sec:case_dfr}

Flexible loads can provide fast frequency support but may also induce flapping when many units switch in an uncoordinated manner.  Each DFR $i$ is modelled as in Section \ref{sec:models} and their switching logic follows a decentralized stochastic policy evaluated every $\Delta t_{\rm ctrl}$, as described by \eqref{eq:logicDFR}.

The WSCC 9-Bus Test System considers three PQ loads of 2.0, 0.9, and 1.0 pu at buses 5, 6, and 8.  The 2.0 pu load at bus 5 is subdivided into three groups (1.1, 0.6, and 0.3 pu) and partially represented by $n=20$ DFRs of 0.01 pu each ($\approx 6\%$ of the total load).  Each DFR measures its local frequency $\omega_{i}(t)$ every $\Delta t_{\rm det}=0.1$\,s using PLL-filtered data.  

To detect flapping in DFRs, we employ the algorithm presented in Section \ref{sec:method} with parameters selected to resolve the characteristic cycling period, while ensuring robustness to noise and real-time operation.  The sampling interval is set to $\Delta t_{\mathrm{det}} = 0.1\text{ s},$ providing ten samples per period.  A moving window of $T_w = 12\text{ s}\quad(N_w = \lceil T_w/\Delta t_{\mathrm{det}}\rceil = 120)$ captures multiple oscillation periods, and a shift of $T_s = 3\text{ s}\quad(N_s = 30)$ balances detection latency against computational load.  Lag bounds corresponding to flapping periods $T_{\min}=0.9$\,s and $T_{\max}=1.1$\,s are converted to $ k_{\min} = 9,\quad k_{\max} = 11$ as per \eqref{eq:lag_bounds_expanded}.  The detection threshold $R_{\mathrm{th}}=0.90$, tolerance $\varepsilon=10^{-3}$, and persistence count $M=4$ are selected based on noise‐level analysis and desired false‐alarm rate.  At each shift, the normalized autocorrelation $\rho_i[k]$ is computed via \eqref{eq:acf_sample_expanded} and \eqref{eq:acf_norm_expanded}, and the metric $r_i^*$ from \eqref{eq:metric_expanded} is evaluated.  The persistence counter $c_i$ and flapping flag $\phi_i$ are then updated following \eqref{eq:counter_expanded} and \eqref{eq:flag_expanded}.  

\begin{figure}[htb]
\centering
\includegraphics[width=0.48\textwidth]{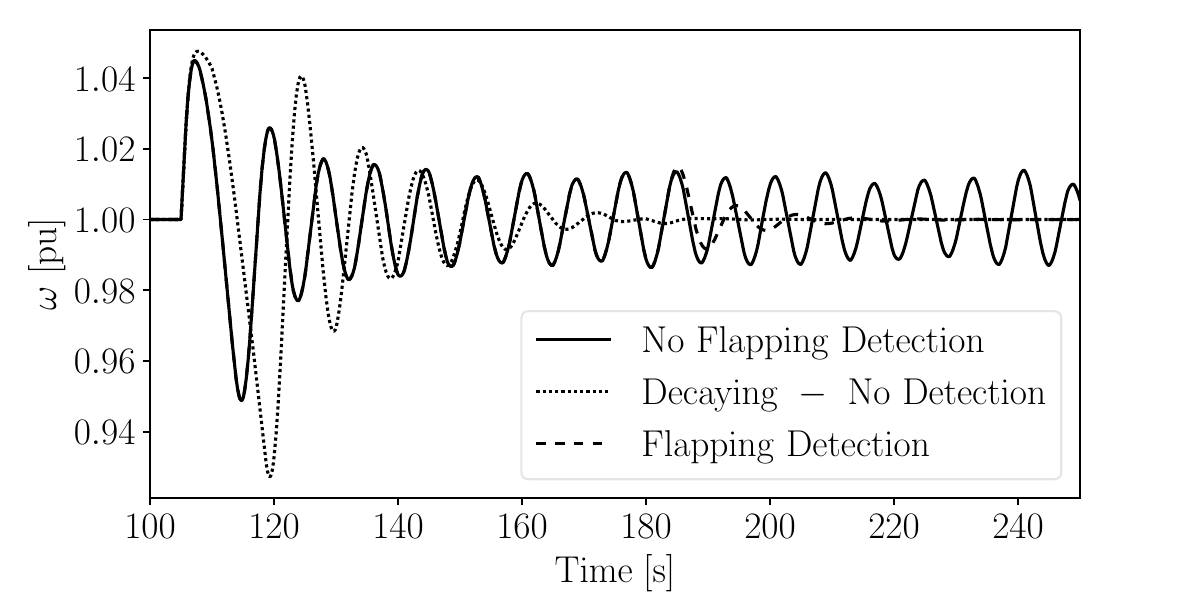}
\caption{DFR flapping detection and flag $\phi_i$.}
\label{fig:flappinginertia1}
\end{figure}

At $t=5$ s, the tripping of the 1.1 pu load initiates a frequency excursion.  Figure \ref{fig:flappinginertia1} shows three scenarios: (i) no control to detect flapping is implemented; (ii) a decaying oscillation occurs, so DFR's flapping flag $\phi_i$ correctly remains zero because it is not a sustained oscillation; and (iii) the flapping controller remains inactive during the damped transient (105–150 s) and switches to one under sustained oscillations ($\approx$180 s).

\begin{figure}[htb]
\centering
\includegraphics[width=0.48\textwidth]{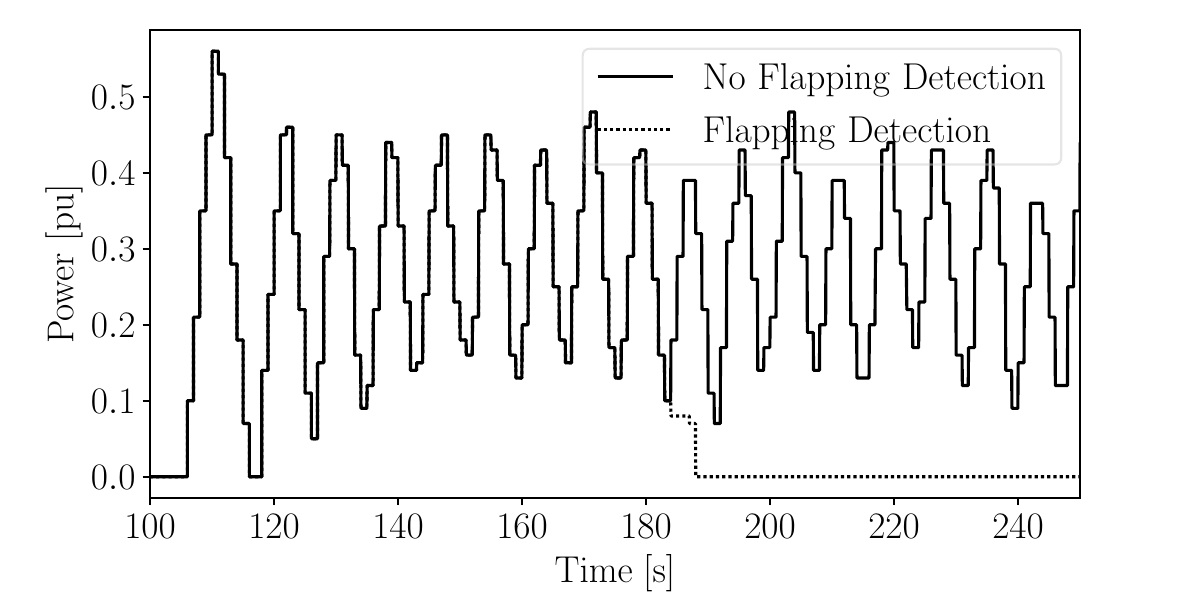}
\caption{Aggregate DFR power injection.}
\label{fig:FDRparticipation}
\end{figure}

Once flapping is detected, DFRs autonomously reduce their switching probability.  Figure \ref{fig:FDRparticipation} illustrates that the aggregate power remains steady during benign oscillations but decreases after detection, restoring system frequency without centralized coordination.  

\begin{figure}[htb]
\centering
\includegraphics[width=0.48\textwidth]{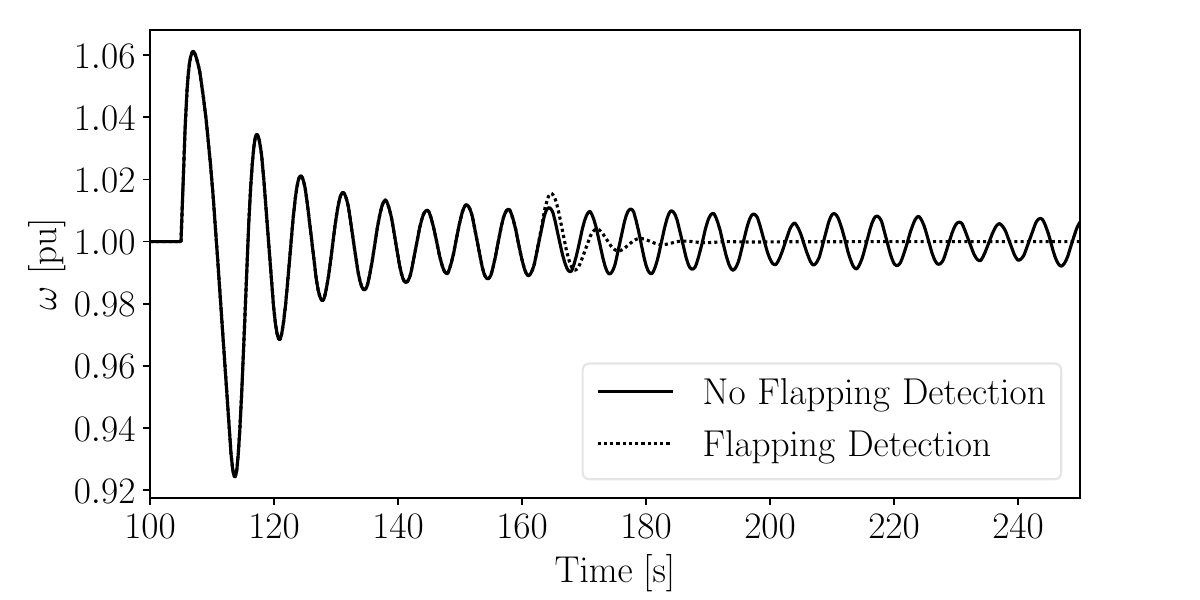}
\caption{Flapping detection under reduced inertia.}
\label{fig:lower_inertia}
\end{figure}

Robustness is tested under reduced system inertia (40\% lower synchronous-machine inertia).  Despite faster and larger oscillations, Fig.~\ref{fig:lower_inertia} shows that the same detector settings successfully identify flapping and trigger mitigation.  This confirms that detection is relative to oscillatory patterns rather than absolute thresholds.  

\begin{figure}[htb]
\centering
\includegraphics[width=0.48\textwidth]{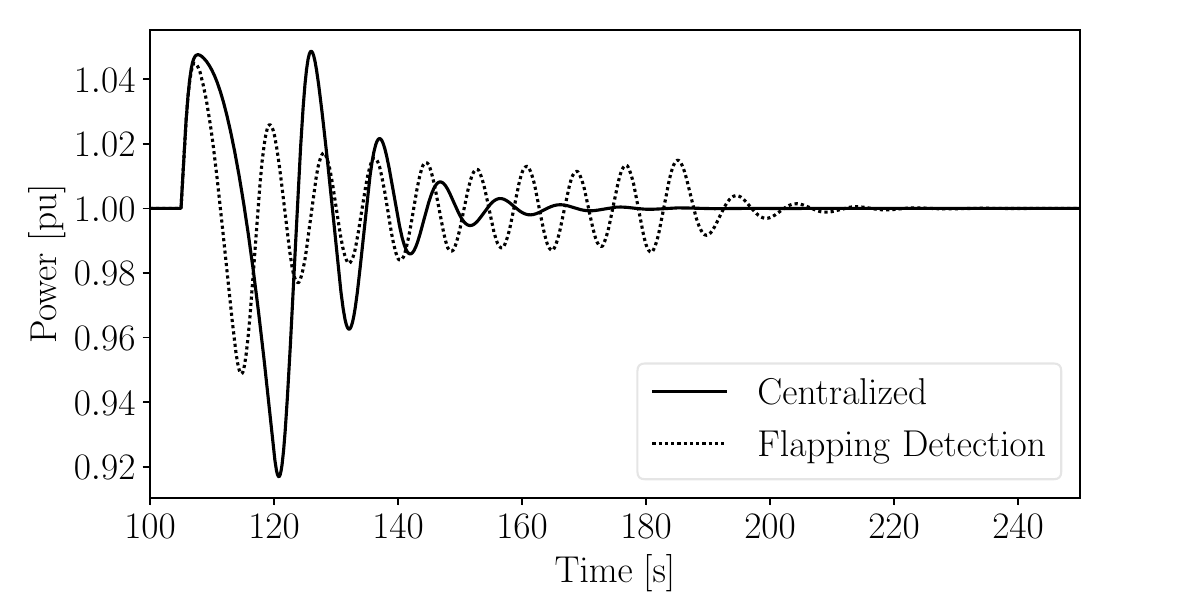}
\caption{Frequency response: decentralized vs.\ centralized tuning.}
\label{fig:comparison}
\end{figure}

Finally, Fig.~\ref{fig:comparison} compares the decentralized scheme against a conservatively tuned centralized controller (only 50\% of the DFRs are operating at the same time).  The centralized strategy suppresses flapping but limits the frequency response, especially at the initial oscillations of the frequency.  In contrast, the decentralized method allows stronger early support while preventing sustained oscillations, at the cost of having more oscillations due to the operation of the DFRs.  %achieving superior overall performance.

\subsection{Under-Load Tap Changer}
\subsubsection{Base Model of the ULTC}

The second case study focuses on analyzing the Under Load Tap Changer Transformer (ULTC), which operates in discrete steps to control the voltage of the downstream bus to which it is connected.

\begin{figure}[ht]
    \centering
    \includegraphics[width=0.45\textwidth]{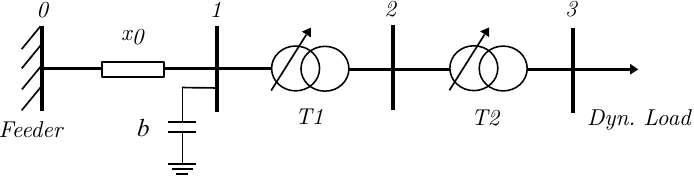}
    \caption{Two ULTC distribution system.}
    \label{fig:ULTC}
\end{figure}

The analysis scenario is a system explained in \cite{Hiskens}, in which two ULTCs are connected in series, as in Fig.~\ref{fig:ULTC}, and when seeking to control the voltage of their respective buses, they have a complementary and sustained operation in time, which causes a phenomenon named ``limit cycle'' or ``cycling.''

This behavior has been the subject of several analyses and publications, but none of them has achieved a mathematical approach to model the instability conditions.

The equation that describes the operation of the ULTC, considering the tap position $m_{k}$ as the state variable, is:
\begin{equation}
m_{k+1}= m_{k} + \Delta m    
\end{equation}
where:
\begin{equation}
    \Delta m =
    \begin{cases}
      +\Delta m & \text{if } v > v^{\max}, \\
      - \Delta m & \text{if } v < v^{\min}, \\
      0 & \text{if } v^{\min}\leq v \leq v^{\max}\ \\
      0 & \text{if} \ m_{k}\ \geq m_{\max}\ \\
      0& \text{if} \ m_{k} \leq m_{\min}
    \end{cases}
\end{equation}
The load is a dynamic exponential recovery load, following:
\begin{align}
        T_p \dot{x}_p &= p_s(v) - p_t(v) - x_p(t) \\
        T_q \dot{x}_q &= q_s(v) - q_t(v) - x_q(t) \\
        p_3 &= p_t(v) - x_p(t) \\
        q_3 &= q_t(v) - x_q(t),    
\end{align} 

As the initial parameters of the system, we have $x_0$ = 0.1 pu, b = 3.33 pu, $\Delta m^1$ = 0.02 pu, $\Delta m^2$ = 0.05 pu, $v_0$ = 1.0 pu.  Initial tap ratios: $m_0^{T1}$ = $m_0^{T2}$= 1.  Delay of the ULTCs: $T_{\text{tap}}^1$ = 53, $T_{\text{tap}}^2$ = 19; $v^{max}$ =1.01 pu, $v^{min}$=0.99 pu.  And regarding the dynamic load, $T_p = 10 \, \text{s}$, $T_q = 5 \, \text{s}$, $p_s = p_0 = 2 \, \text{pu}$, $q_s = q_0 = 0.5 \, \text{pu}$, $p_t = p_0 v^2$, and $q_t = q_0 v$.\\

\begin{figure}[ht]
    \centering
    \includegraphics[width=0.5\textwidth]{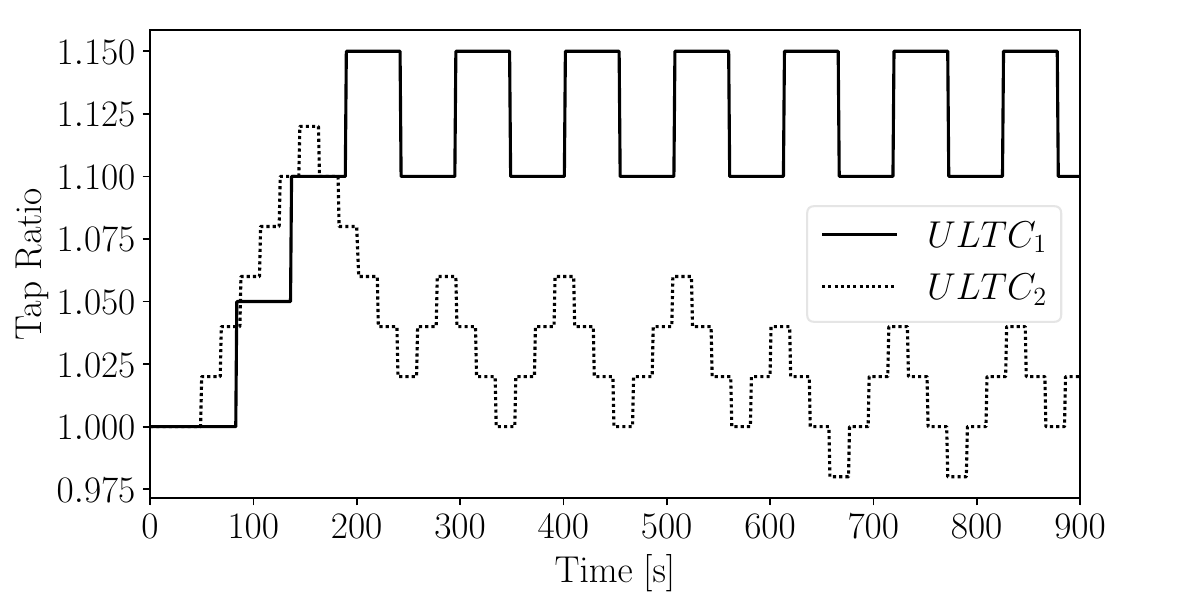}
    \caption{Time response of the two ULTC}
    \label{fig:ULTCcascade}
\end{figure}

As observed in Fig.~\ref{fig:ULTCcascade}, the responses of each ULTC correspond to an undesirable condition.
    
Then, we apply the detection algorithm with:
\[
\Delta t_{\mathrm{det}} = 1\,\text{s},\;
T_w = 300\,\text{s}\;(N_w=300),\;
T_s = 30\,\text{s}\;(N_s=30).
\]
Oscillation periods in $[T_{\min},T_{\max}]=[80,120]\,\text{s}$ yield lag bounds $k_{\min}=80$, $k_{\max}=120$.  We maintain a circular buffer of length $N_w + k_{\max}$ for the controlled voltage $v_{\mathrm{con}}[n]$.  Detection parameters are set as $R_{\mathrm{th}}=0.35$, $\varepsilon=10^{-3}$.  At each shift, we compute $\rho[k]$ evaluate $r^*$, and update $c,\phi$.  This configuration reliably flags slow limit-cycle oscillations while rejecting faster voltage transients.

The detector uses $\Delta t_{\rm det}=1$\,s, $T_w=300$\,s, $T_s=30$\,s, lag bounds $k_{\min}=80$, $k_{\max}=120$, and threshold $R_{\mathrm{th}}=0.35$.

\begin{figure}[htb]
  \centering
  \includegraphics[width=0.48\textwidth]{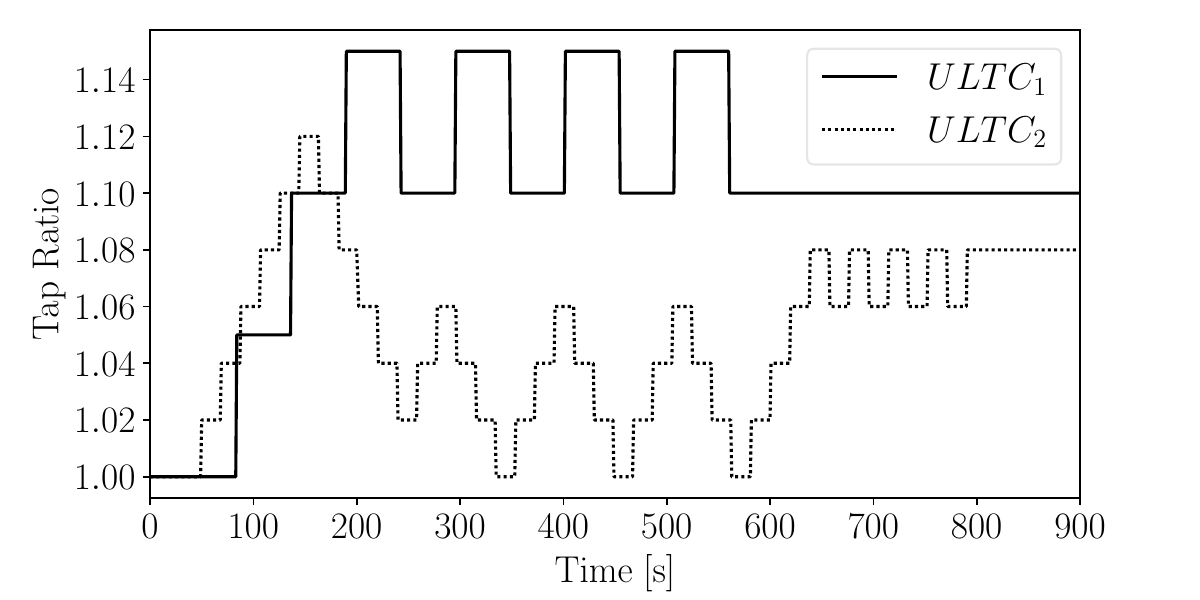}
  \caption{Upstream ULTC flapping flag $\phi$ and blocking action.}
  \label{fig:FlappingSoloved}
\end{figure}

Figure \ref{fig:FlappingSoloved} shows the tap ratio of the ULTCs.  After four consecutive windows of high normalized autocorrelation ($r^*>R_{\mathrm{th}}$) flapping is detected, and triggers the blocking action.  

In our coordinated scheme, the upstream ULTC is assigned a lower persistence count $M = 4$, reflecting its larger tap step size and greater influence on voltage stability, while the downstream ULTC uses a higher $M$ to tolerate benign fluctuations.  When the upstream device’s flag $\phi$ transitions to 1 at around $t=550$\,s, its tap change $\Delta m$ is set to zero, effectively freezing its operation and allowing the downstream transformer to re-establish voltage control without flapping.  If residual oscillations persist beyond a subsequent detection window, the downstream ULTC also freezes, as seen at $t\approx800$ s.

\begin{figure}[htb]
  \centering
  \includegraphics[width=0.48\textwidth]{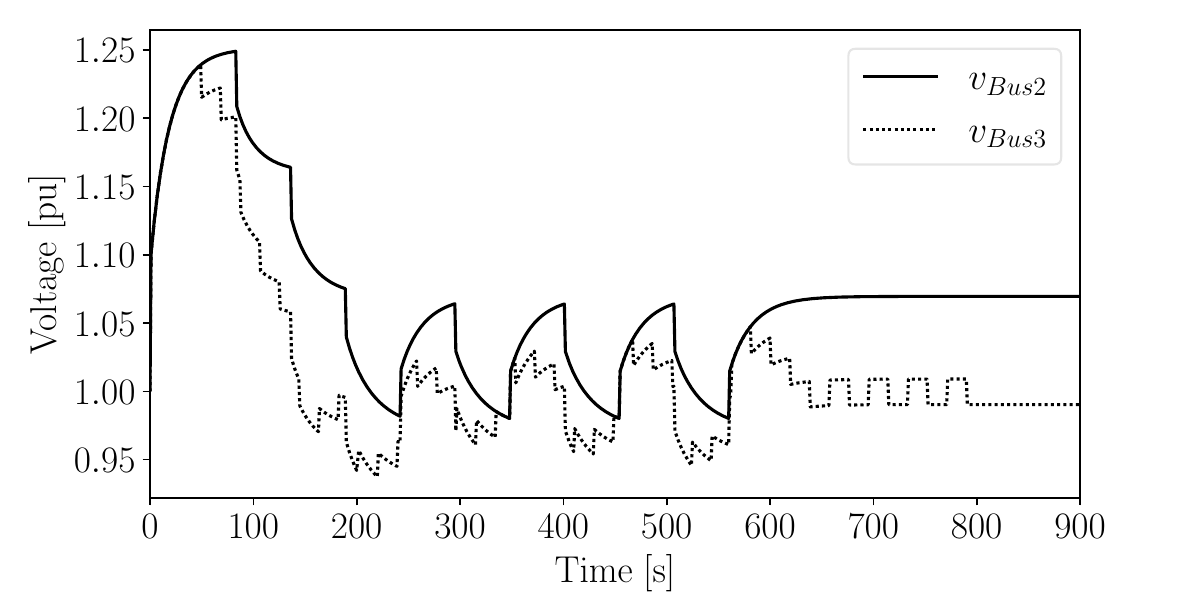}
  \caption{Controlled voltages at transformer terminals before and after flapping mitigation.}
  \label{fig:Voltages}
\end{figure}

Figure \ref{fig:Voltages} compares the terminal voltages of the upstream and downstream ULTCs before and after mitigation.  Prior to detection, both transformers cycle between their tap limits, producing voltage oscillations of approximately $\pm 5\%$ around the 1.0 pu setpoint.

As shown in Fig.~\ref{fig:Voltages}, once the upstream ULTC is blocked, the downstream transformer reduces voltage deviations to within ±0.2\% of the nominal value.  After the downstream freeze at $t\approx800$\,s, the voltages remain steady and within an acceptable boundary.

This case confirms that our decentralized, autocorrelation-based method reliably identifies and stops ULTC cycling without compromising its objective of voltage regulation.

\subsection{Limit Cycle induced by AVR-Weak Grid Coupling}
\label{sec:case_avr}

While the previous cases focus primarily on discrete elements such as DFRs and ULTCs, devices with continuous-nature operation, including excitation systems and Automatic Voltage Regulators (AVRs), can also induce sustained oscillatory behaviour.  Such conditions, frequently referred to as limit cycles, exhibit analogous harmful behaviour to flapping.  This section explores the application of the autocorrelation-based detection methodology to continuous-operation devices.

The IEEE 14-bus system consists of 14 buses, 5 synchronous generators, and 11 loads.  Each generator is equipped with an AVR of Type-1 as described in \cite{milano_power_2010}.

First, as proposed in the Section \ref{sec:method}, we establish the detection method based on the tracking of the voltages of the buses of the system.
To induce sustained voltage oscillations, the AVR amplifier gains, $K_{a, i}$, are increased within a range of 240 to 300, intentionally destabilizing their feedback loops.  The only reason why $K_{a, i}$ differs between AVRs is to have a better visual perception when observing their status over time.

We need that each AVR autonomously assess its contribution to the detected oscillations and react accordingly.  For this coordination, as explained in Section \ref{sec:models}, we identify the variable that causes the major impact on the controller compared to the interest value.  From the model of the AVRs, we can state that relevant variables as the excitation field of the synchronous machine, the feedback signal of the AVR, among others, have a direct impact on the observed value (voltage).  For having access to the measurement of the controller, we use the signal related to the feedback signal of the AVR.  
% , $v_{r, 2}$.  
Then, we calculate the cumulative Teager Operator of this variable.

\begin{figure}[htb]
  \centering
  \includegraphics[width=0.49\textwidth]{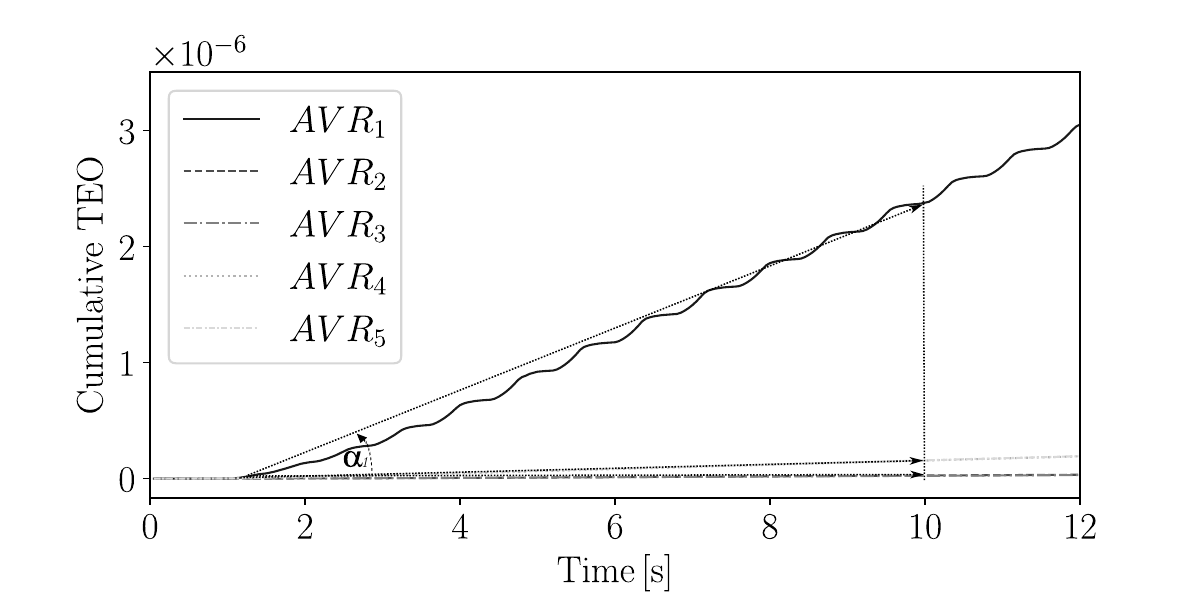}
  \caption{Teager Operator of the transient feedback state}
  \label{fig:TEO}
\end{figure}

Figure \ref{fig:TEO} illustrates the TEO's trajectories, highlighting $AVR_1$ as the primary contributor, followed by $AVR_4$ and $AVR_5$.  Conversely, $AVR_2$ and $AVR_3$ exhibit negligible impact.  This distinction is crucial for prioritizing mitigation actions.

From the TEO's representation, the contribution to the disturbance is calculated as the angle $\alpha_i$, normalized by $\alpha_{\text{max}} = 90^\circ$ applying (\ref{eq:alfa}).  This normalized value is used as a probabilistic modulation factor: AVRs with a larger estimated impact on the oscillation are more likely to switch.

In principle, if a suitable communication infrastructure were available, a centralized operator could compare the impact metrics from all AVRs, coordinate their responses, and prescribe the optimal corrective actions.  However, to preserve decentralization and avoid the delays and complexity of central control, we propose a stochastic local decision‐making scheme, ruled by (\ref{eq:TEOoperation}).  Under this scheme, each AVR evaluates its operation logic at every control interval $M_{\rm ctrl}$ defining $K_{a,i}$.  The value $K_{a,i}^{\text{safe}}$ denotes a reduced AVR gain, chosen to be sufficient for maintaining stability while preserving basic voltage regulation capability, whereas $K_{a,i}^{\text{init}}$ corresponds to the original operating gain.

Applying the method in the IEEE 14-Bus Test System by injecting a perturbation at $t=1s$, which triggers the oscillatory behaviour of the voltage.  An increasing amplitude of the oscillations is observed in the first seconds, but then it stays as a sustained oscillation.

\begin{figure}[htb]
  \centering
  \includegraphics[width=0.48\textwidth]{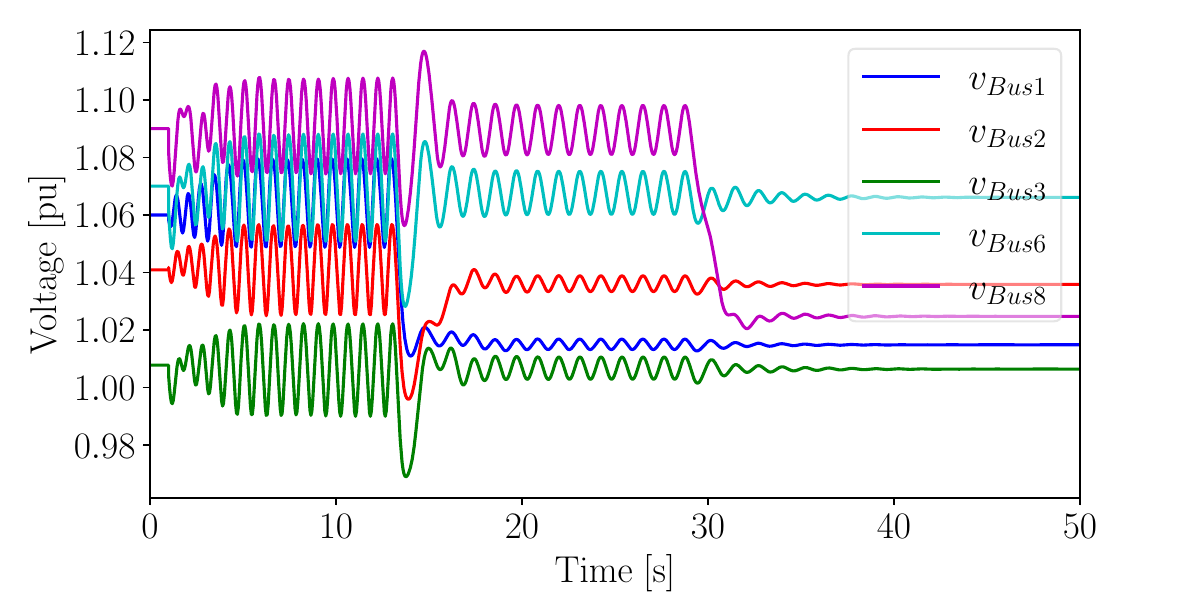}
  \caption{Oscillatory response of the voltages}
  \label{fig:Voltages14Bus}
\end{figure}

\begin{figure}[htb]
  \centering
  \includegraphics[width=0.48\textwidth]{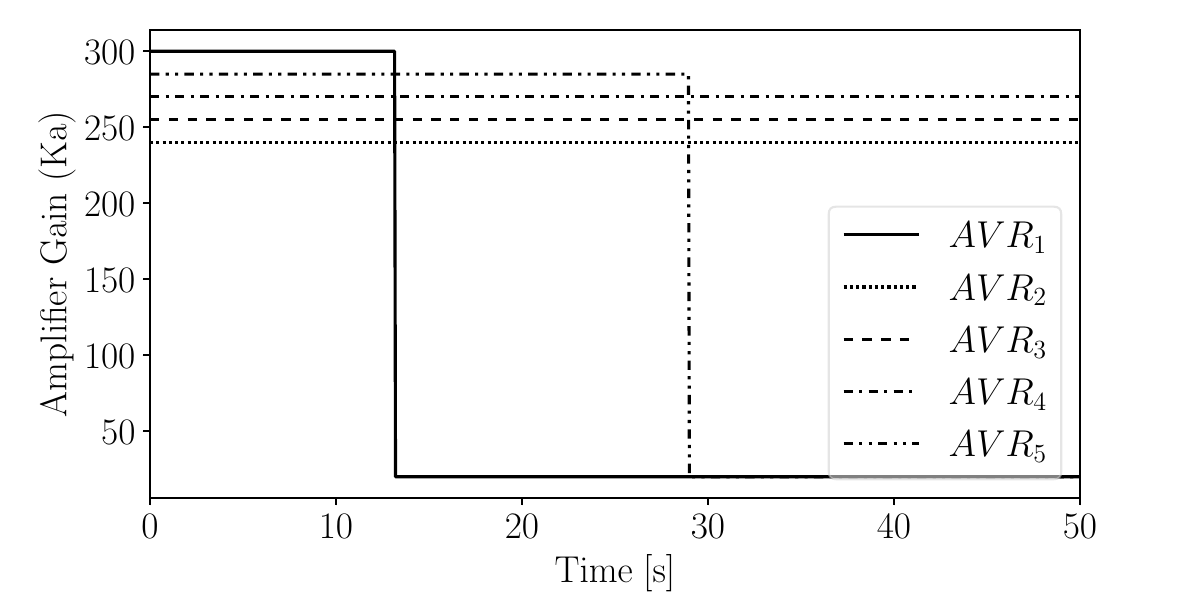}
  \caption{Amplifier gain $K_{a}$ behaviour over time}
  \label{fig:GainAVR}
\end{figure}

Following the logic of the method, while the controllers to detect flapping are running, in parallel, the cumulative TEO is being calculated, and the limits of (\ref{eq:alfa}).  Once it is the moment to operate (the flapping is detected), the random number is generated, and will define if the $i_{th}$ AVR will move its gain $Ka$, according to (\ref{eq:TEOoperation}), as can be tracked in Fig.\ref{fig:GainAVR}.  Figure \ref{fig:Voltages14Bus} shows that $\rm AVR_1$, which has the highest $\tilde{\alpha}_i$, indeed operates first; however, its operation is insufficient to control the voltage oscillations.  Therefore, the other AVRs continue to generate random values and evaluate their operation.  Around $t=30s$, $\rm AVR_5$ reacts and changes its gain.  With this second action, the system has accomplished enough actions to solve the oscillations of the voltages, so no more AVRs should respond.

\begin{figure}[htb]
  \centering
  \includegraphics[width=0.48\textwidth]{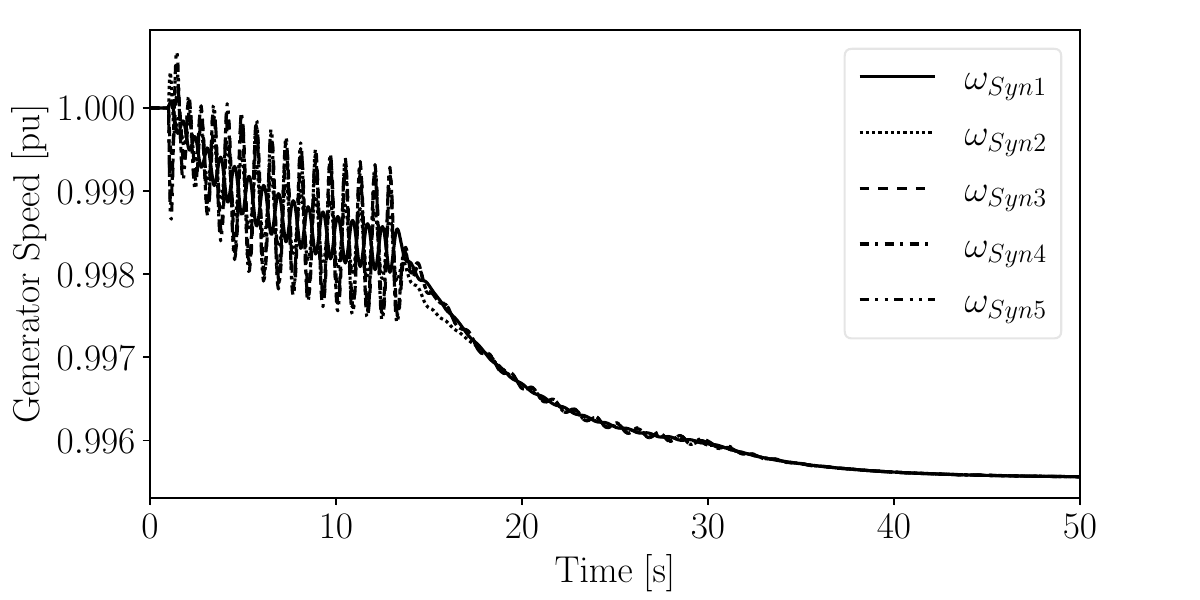}
  \caption{Speed of the machines}
  \label{fig:speedMachines}
\end{figure}

Finally, Fig.~\ref{fig:speedMachines} shows the synchronous generator frequencies, where it can be observed how the machines react to the voltage oscillations of the system.  At this point, it is essential to emphasize the importance of identifying and utilizing the correct variable to calculate the TEO and calibrate the actions of the controllers.  Because, despite the influence of other variables on the system's voltage, such as reactive power injection, it can cause undesired results.

\section{Remarks and Practical Considerations} 
\label{sec:comments}

The effectiveness of the proposed autocorrelation-based detection critically depends on both signal quality and appropriate parameter tuning.  In practical implementations, preprocessing techniques, such as low-pass filtering or spike removal, can improve the reliability of the autocorrelation coefficients $\rho[k]$ in noisy environments.  Care must be taken, however, to preserve the characteristic frequency content of flapping phenomena, since excessive filtering may hide oscillations or misclassify them as slow drifts.

Parameter tuning should reflect the dynamic characteristics of the targeted devices.  For DFRs, characterized by rapid switching on the order of seconds, short analysis windows (e.g., $T_w \sim 10$ s) and narrow lag bands ($T_{\text{min}} \approx 1$ s) are effective.  Conversely, ULTCs, which operate over longer periods, benefit from extended windows (e.g., $T_w \sim 300$ s) and wider lag intervals (e.g., $T_{\text{min}} \approx 80$ s).  Operators should estimate expected oscillation periods in advance and calibrate parameters such as $T_w$, $k_{\text{min}}$, $k_{\text{max}}$, $R_{\text{th}}$, and the persistence threshold $M$ accordingly.

While simpler detection approaches exist, such as counting discrete device operations, these can lead to false positive detection in case of poorly damped oscillations, stochastic switching, or changes in the device population.  False positive detections unnecessarily degrade the dynamic performance of the system.  By relying instead on continuous system-level measurements, the proposed autocorrelation-based method provides robust detection that is less sensitive to such variations.  The methodology can also be extended to continuous-operation devices, such as AVRs, as discussed in Section \ref{sec:case_avr}, enabling a unified approach to oscillation detection across device types.

\section{Conclusion}
\label{sec:conclusion}

This work presents a decentralized methodology based on autocorrelation analysis to identify and address flapping phenomena in power systems.  In the proposed approach, individual devices detect sustained oscillations by means of local measurements, and respond autonomously, without reliance on centralized coordination or extensive communication infrastructure.

The proposed framework supports diverse device types, including DFRs, ULTCs, and continuous-operation AVRs, through parameter sets tailored to their specific dynamic characteristics.  This ensures reliable discrimination between harmful persistent oscillations and benign transients, even under variable operating conditions.  The method’s scalability and robustness have been validated through simulations, showing consistent improvements in system stability and control effectiveness.

Future work will explore adaptive threshold tuning, integration with hybrid centralized–decentralized control architectures, and broader application of the methodology to other stability phenomena in low-inertia systems.

% ======================================================================
%\bibliographystyle{IEEEtran}
%\bibliography{references}


\end{document}